\documentclass[twocolumn]{aastex63}

\usepackage{subfigure}

\usepackage{mathtools}

\DeclarePairedDelimiter{\nint}\lfloor\rceil

\received{January 11, 2021}
\revised{February 23, 2021}
\accepted{February 25, 2021}
\submitjournal{AJ}

\shorttitle{Tails: Chasing Comets with ZTF and Deep Learning}
\shortauthors{Duev et al.}
\graphicspath{{./}{figures/}}

\begin{document}

\title{Tails: Chasing Comets with the Zwicky Transient Facility and Deep Learning}

\correspondingauthor{Dmitry A. Duev}
\email{duev@caltech.edu}

\author[0000-0001-5060-8733]{Dmitry A. Duev}
\affiliation{Division of Physics, Mathematics, and Astronomy, California Institute of Technology, Pasadena, CA 91125, USA}

\author[0000-0002-4950-6323]{Bryce T. Bolin}
\affiliation{Division of Physics, Mathematics, and Astronomy, California Institute of Technology, Pasadena, CA 91125, USA}\affiliation{IPAC, California Institute of Technology, 1200 E. California Blvd, Pasadena, CA 91125, USA}

\author[0000-0002-3168-0139]{Matthew J. Graham}
\affiliation{Division of Physics, Mathematics, and Astronomy, California Institute of Technology, Pasadena, CA 91125, USA}

\author[0000-0002-6702-7676]{Michael S. P. Kelley}
\affiliation{Department of Astronomy, University of Maryland, College Park, MD 20742, USA}

\author[0000-0003-2242-0244]{Ashish Mahabal}
\affiliation{Division of Physics, Mathematics, and Astronomy, California Institute of Technology, Pasadena, CA 91125, USA}
\affiliation{Center for Data Driven Discovery, California Institute of Technology, Pasadena, CA 91125, USA}



\author[0000-0001-8018-5348]{Eric C. Bellm}
\affiliation{DIRAC Institute, Department of Astronomy, University of Washington, 3910 15th Avenue NE, Seattle, WA 98195, USA}

\author[0000-0002-8262-2924]{Michael W. Coughlin}
\affiliation{School of Physics and Astronomy, University of Minnesota, Minneapolis, Minnesota 55455, USA}

\author[0000-0002-5884-7867]{Richard Dekany}
\affiliation{Caltech Optical Observatories, California Institute of Technology, Pasadena, CA 91125, USA}

\author[0000-0003-3367-3415]{George Helou}
\affiliation{IPAC, California Institute of Technology, 1200 E. California Blvd, Pasadena, CA 91125, USA}

\author[0000-0001-5390-8563]{Shrinivas R. Kulkarni}
\affiliation{Division of Physics, Mathematics, and Astronomy, California Institute of Technology, Pasadena, CA 91125, USA}

\author[0000-0002-8532-9395]{Frank J. Masci}
\affiliation{IPAC, California Institute of Technology, 1200 E. California Blvd, Pasadena, CA 91125, USA}

\author[0000-0002-8850-3627]{Thomas A. Prince}
\affiliation{Division of Physics, Mathematics, and Astronomy, California Institute of Technology, Pasadena, CA 91125, USA}

\author[0000-0002-0387-370X]{Reed Riddle}
\affiliation{Caltech Optical Observatories, California Institute of Technology, Pasadena, CA 91125, USA}

\author[0000-0001-6753-1488]{Maayane T. Soumagnac}
\affiliation{Lawrence Berkeley National Laboratory, 1 Cyclotron Road, Berkeley, CA 94720, USA}
\affiliation{Department of Particle Physics and Astrophysics, Weizmann Institute of Science, Rehovot 76100, Israel}

\author[0000-0001-9276-1891]{St\'{e}fan J. van der Walt}
\affiliation{Berkeley Institute for Data Science, University of California, Berkeley, Berkeley, CA 94720, USA}

\begin{abstract}

We present Tails, an open-source deep-learning framework for the identification and localization of comets in the image data of the Zwicky Transient Facility (ZTF), a robotic optical time-domain survey currently in operation at the Palomar Observatory in California, USA. 
Tails employs a custom EfficientDet-based architecture and is capable of finding comets in single images in near real time, rather than requiring multiple epochs as with traditional methods.
The system achieves state-of-the-art performance with  99\% recall, 0.01\% false positive rate, and 1-2 pixel root mean square error in the predicted position. 
We report the initial results of the Tails efficiency evaluation in a production setting on the data of the ZTF Twilight survey, including the first AI-assisted discovery of a comet (C/2020 T2) and the recovery of a comet (P/2016 J3 = P/2021 A3).

\end{abstract}


\section{Introduction} \label{sec:intro}
Comets have mesmerized humans for millennia, frequently offering, arguably, some of the most spectacular sights in the night sky. Containing the original materials from when the Solar System first formed, comets provide a unique insight into the distant past of our Solar System. The recent discovery of the first interstellar comet 2I/Borisov by amateur astronomer Gennadiy Borisov predictably sparked much excitement and enthusiasm among astronomers and the general public alike \citep[e.g.,][]{Bolin2020AJ, 2019ApJ...885L...9F, 2020NatAs...4...53G}. Such objects could potentially provide important information on the formation of other stellar systems. It is a very exciting time to look for comets: the large-scale time-domain surveys that are currently in operation, such as the ZTF \citep{2019PASP..131a8002B, 2019PASP..131g8001G}, Pan-STARRS \citep{2016arXiv161205560C}, or ATLAS \citep{2018PASP..130f4505T}, and  the upcoming ones such as BlackGEM \citep{2016SPIE.9906E..64B} and Vera Rubin Observatory / LSST \citep{ivezic2008lsst} offer the richest data sets ever available to mine for comets.

Traditional comet detection algorithms rely on multiple observations of cometary objects that are linked together and used to fit an orbital solution. To the best of our knowledge, the previous attempts to take the comet's morphology in the optical image data into consideration in the detection algorithms have not led to reliable and robust results.

In this work, we present Tails -- a state-of-the-art deep-learning-based system for the identification and localization of comets in the image data of ZTF. Tails employs an EfficientDet-based architecture \citep{2019arXiv191109070T} and is thus capable of finding comets in single images in near real time, rather than requiring multiple epochs as with traditional methods.

The Tails' code is open-source and can be found in the \href{https://github.com/dmitryduev/tails}{``dmitryduev/tails''} repository on GitHub. The version of the code aligned with this publication is archived on Zenodo at \href{https://doi.org/10.5281/zenodo.4563226
}{10.5281/zenodo.4563226}.

\subsection{The Zwicky Transient Facility}

The Zwicky Transient Facility (ZTF)\footnote{\url{https://ztf.caltech.edu}} is a state-of-the-art robotic time-domain sky survey capable of visiting the entire visible sky north of $-30^\circ$ declination every night. ZTF observes the sky in the $g$, $r$, and $i$ bands at different cadences depending on the scientific program and sky region \citep{2019PASP..131a8002B, 2019PASP..131g8001G}. The 576 megapixel camera with a 47 deg$^2$ field of view, installed on the Samuel Oschin 48-inch (1.2-m) Schmidt Telescope, can scan more than 3750 deg$^2$ per hour, to a $5\sigma$ detection limit of 20.7 mag in the $r$ band with a 30-second exposure during new moon \citep{2019PASP..131a8003M, 2020PASP..132c8001D}.

The ZTF Partnership has been running a specialized survey, the Twilight Survey (ZTF-TS) that operates at Solar elongations down to 35 degrees with an $r$-band limiting magnitude of 19.5 \citep{2020AJ....159...70Y, 2019PASP..131f8003B}. ZTF-TS has so far resulted in the discovery of a number of Atira asteroids (orbits interior to the Earth's) as well as the first inner-Venus object, 2020 AV2 \citep{2020arXiv200904125I}. Motivated by the success, ZTF-TS will be expanded in Phase II of the project, which commenced in December 2020.

Comets become more easily detectable when close to the Sun as they become brighter and start exhibiting more pronounced coma and tails. Furthermore, it has been shown that the most detectable direction of approach of an interstellar object is from directly behind the Sun because of observational selection effects \citep{Jedicke2016} and the fact that this direction has a greater cross section for asteroids to bend around and pass into the visibility volume \citep{Engelhardt2017, 2018ApJ...855L..10D}.

Tails automates the search for comets with detectable morphology. While trained and evaluated on a large corpus of ZTF data, in this work we focus on Tails' performance when applied to the ZTF-TS data.

\section{Tails: a deep learning framework for the identification and localization of comets} \label{sec:tails}

Deep learning (DL) is a subset of machine learning that employs artificial many-layer neural networks \citep{McCulloch1943}. DL systems are able to discover, in a highly automated manner, efficient representations of the data, simplifying the task of finding the meaningful sought-after patterns in them. We refer the reader to a brilliant introduction into DL given in \citet{Geron}.

DL systems often reach near-optimal performance for a given task and are able to learn even very complicated, highly non-linear mappings between the input and output spaces. The art of building applied DL systems involves two major challenges: finding a suitable network architecture and, more importantly, constructing a large, labeled, representative data set for the network training. In the case of comet detection, the training set must reflect the possible variations across different seeing conditions, filters, sky location, CCDs, and include data artifacts caused by, for example, cross-talk or telescope reflections.

\subsection{Data set}

\begin{figure}
    \centering
    \includegraphics[width=0.47\textwidth]{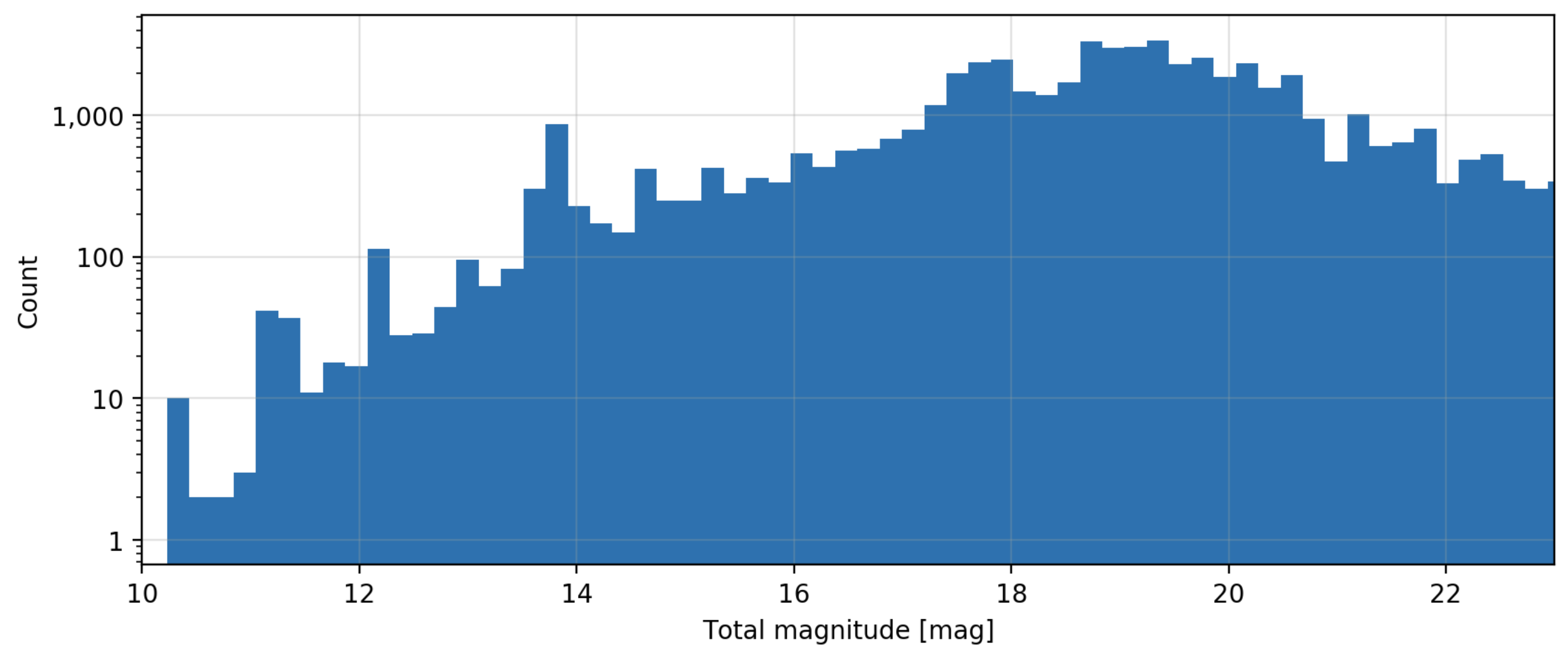}
    \caption{Distribution of over 60,000 individual observations of comets as a function of the predicted total magnitude (as reported by JPL Horizons) used in the seed sample.}
    \label{fig:seed-hist}
\end{figure}

\begin{figure*}
    \centering
        \subfigure[114P/Wiseman–Skiff observed on 2019/10/19 ]{\includegraphics[width=0.48\textwidth]{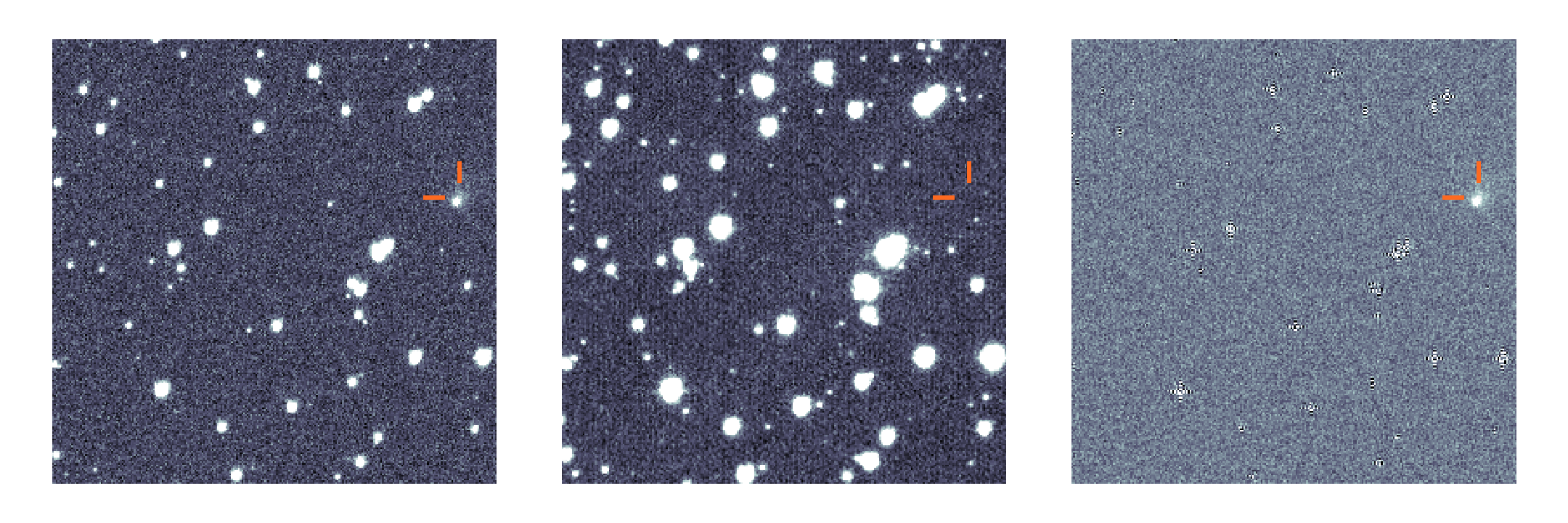}}\quad
        \subfigure[29P/Schwassmann–Wachmann observed on 2019/10/19]{\includegraphics[width=0.48\textwidth]{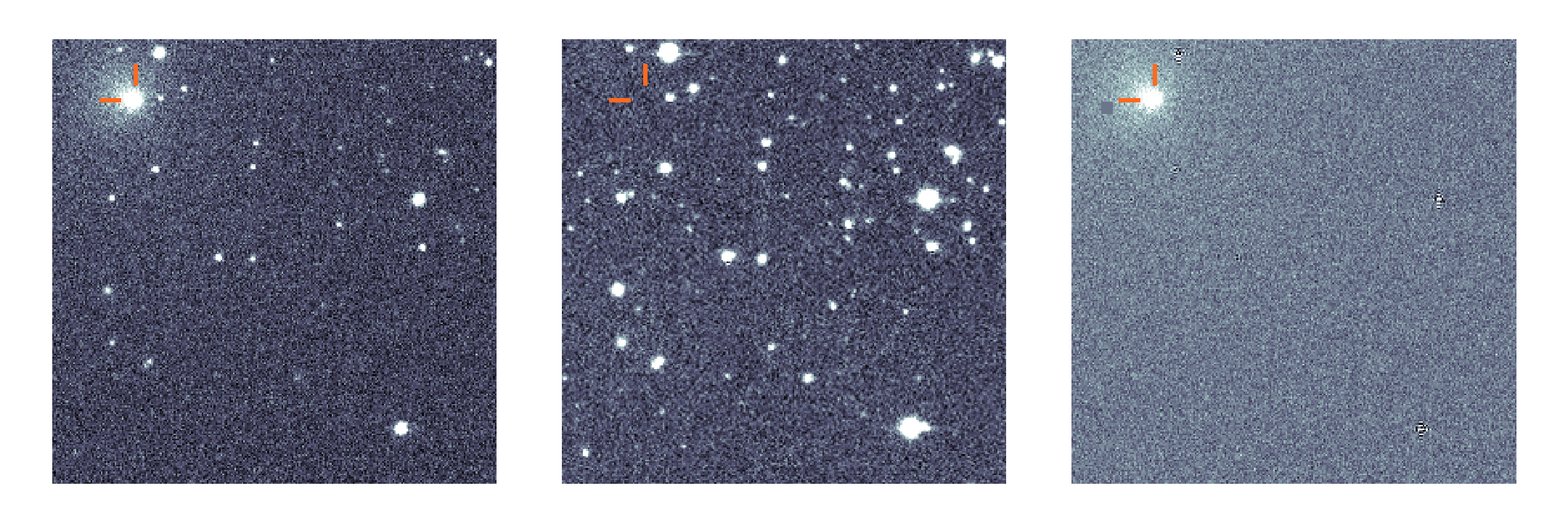}}\quad
        \subfigure[C/2019 D1 (Flewelling) observed on 2019/07/22]{\includegraphics[width=0.48\textwidth]{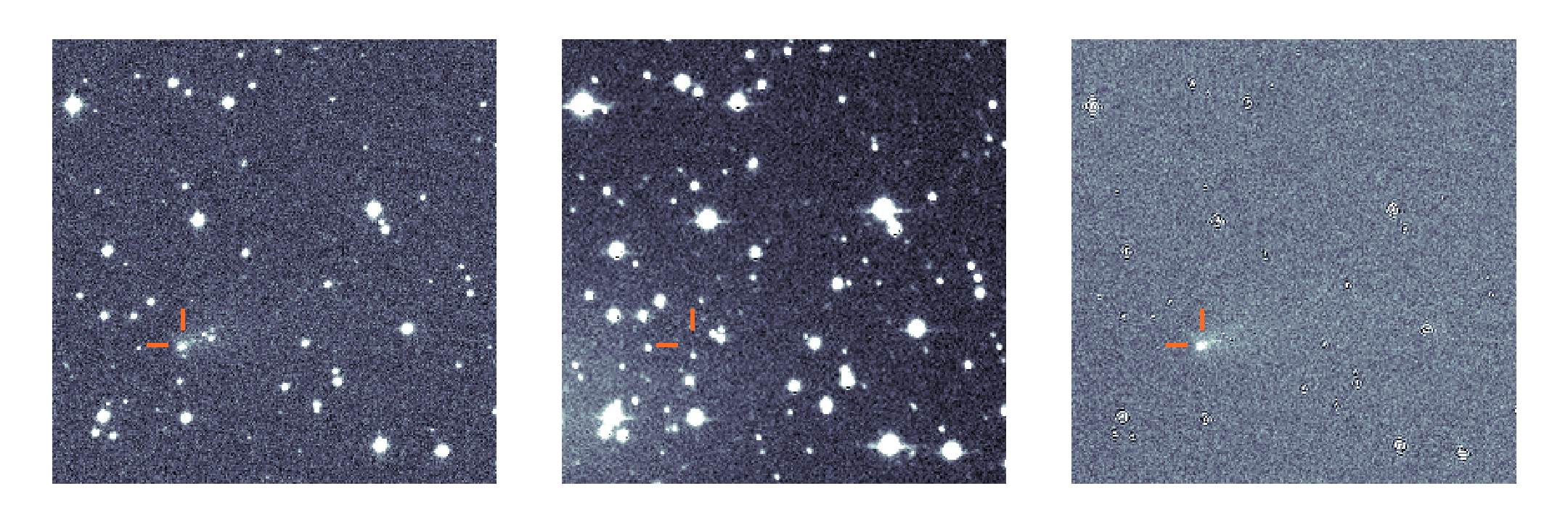}}\quad
        \subfigure[260P/McNaught observed on 2019/10/19]{\includegraphics[width=0.48\textwidth]{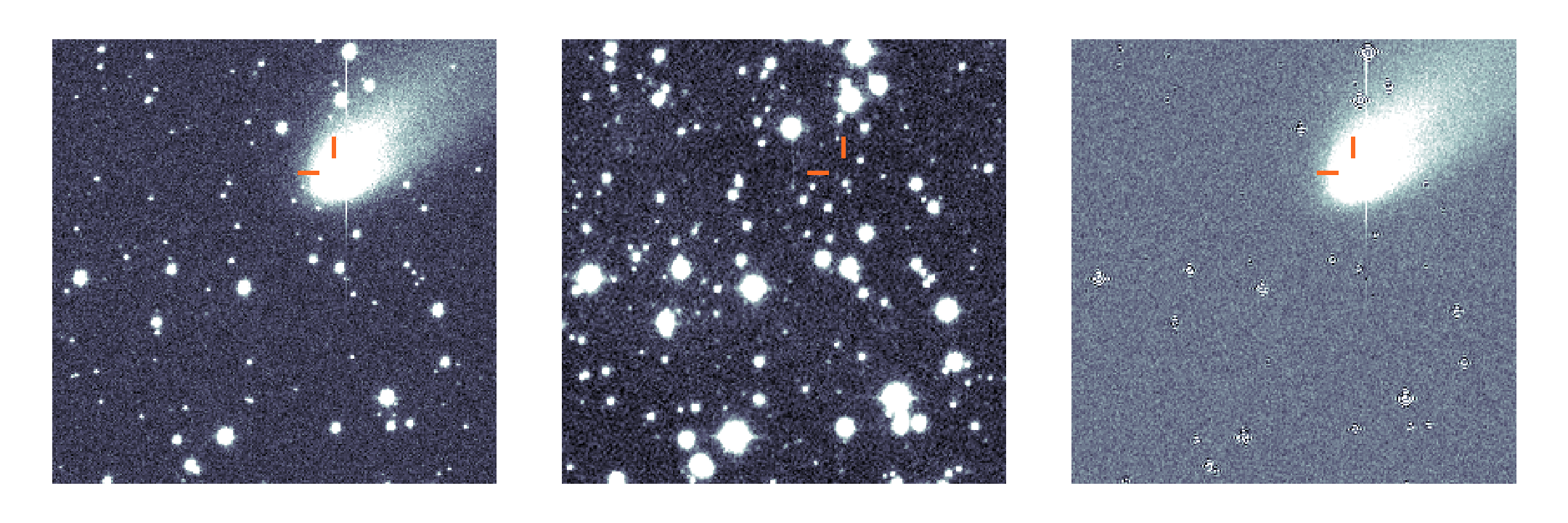}}\quad
    \caption{Comet examples from the training set. The first image in a triplet is the epochal science image, the second -- the reference image, the third -- the ZOGY difference image.}
    \label{fig:comet-examples-training}
\end{figure*}

To build a seed sample for labeling, we first identified all potential observations of known comets conducted with ZTF from March 5, 2018 - March 4, 2020, based on their predicted position and brightness. The code for accomplishing that is based on the Python libraries pypride \citep{2016A&A...593A..34D} and solarsyslib \citep{2018AJ....155...32J} and uses the comet ephemerides obtained from the Minor Planet Center (MPC)\footnote{\url{https://minorplanetcenter.net/iau/MPCORB/CometEls.txt}} for a coarse search, followed by a JPL Horizons\footnote{\url{https://ssd.jpl.nasa.gov/horizons.cgi}} \citep{1996DPS....28.2504G} query for precision.

To provide more contextual information, epochal image data are supplemented by properly aligned reference images of the corresponding patches of sky and difference (epochal minus reference) images generated with the ZOGY algorithm \citep{2016ApJ...830...27Z}, all produced by the ZTF Science Data System at Caltech's IPAC \citep{2019PASP..131a8003M}. Finally, we generate image triplet cutouts of size 256 by 256 pixel, which in angular measure translates into $4\arcmin.3$ by $4\arcmin.3$ at ZTF's pixel scale of $1\arcsec.01 / pix$.

We selected over 60,000 individual observations with the total comet magnitude ranging from 10 to 23 (as reported by JPL Horizons; see Fig. \ref{fig:seed-hist}), out of which about 20,000 were sourced for manual annotation. This resulted in an initial sample of 3,000 examples with identifiable morphology. 

We also compiled a set of approximately 20,000 negative examples consisting of point-like cometary detections, patches of sky with no identified transient or variable sources, CCD-edge cases, and a wide range of real (point-source) transient and bogus (e.g. artifacts due to bright stars, optical ghosts and ``dementors'') samples from the Braai data set \citep{2019MNRAS.489.3582D}.

To expand the data set, we then assembled a standard ResNet-based \citep{2015arXiv151203385H} classifier for comet identification. With this basic classification model, we ran several rounds of an active-learning-like procedure, where we would first train the classifier, evaluate it on the whole data set, sample both confident predictions and the cases close to the classifier's decision boundary, manually inspect and label those examples and add them to the training set. Roughly 2,000 positive and 2,000 negative examples were added to the training set via this method.

The resulting training data set contains about 5,000 positive and 22,000 negative examples (see Fig. \ref{fig:comet-examples-training}). Each triplet in the set has been assigned a label $[p_c, x, y]$, where $p_c$ marks the presence of a comet in the image and $x, y \in [0, 1]$ is the relative positions of the comet's ``center of mass'', as reported by JPL Horizons. For positive examples this translates into $[1, x_{JPL}, y_{JPL}]$, for negative ones -- $[0, ?, ?]$, where question marks mean that these do not affect the loss in this case.

\subsection{Deep neural network architecture and training}

\begin{figure*}
    \centering
    \includegraphics[width=0.99\textwidth]{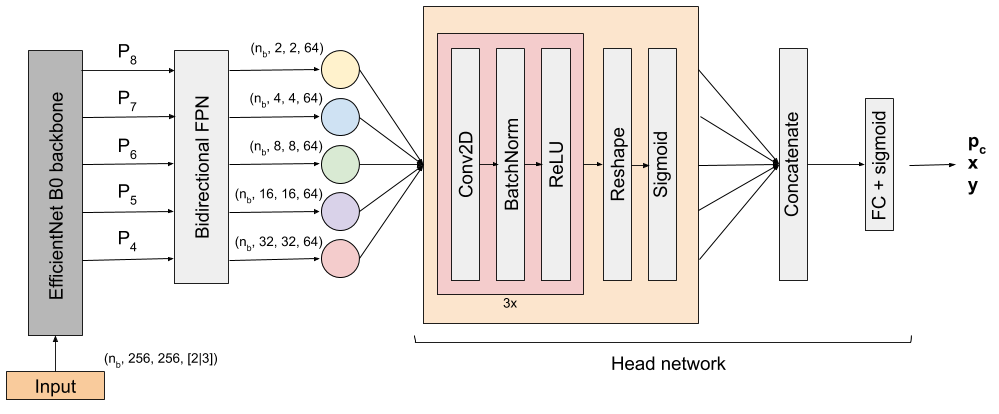}
    \caption{Tails architecture: a custom EfficientDet D0-based network \citep{2019arXiv191109070T}. A batch of duplet or triplet image stacks of size $(n_b, 256, 256, [2|3])$, correspondingly, is passed through an EfficientNet B0 backbone, where $n_b$ is the number of stacks in the batch. The extracted features from the last five blocks/levels of the backbone network are passed through a bidirectional feature-pyramid network (BiFPN). The resulting five output tensors denoted in colored circles are fed into the head network, which outputs the probability of the image containing a comet $p_c$ and its predicted relative position $(x, y)$.}
    \label{fig:architecture}
\end{figure*}

Tails adopts a custom architecture (see Fig. \ref{fig:architecture}) based on EfficientDet D0 \citep{2019arXiv191109070T}, a variant of a state-of-the-art architecture designed for object detection - a computer vision technique for the identification and location of objects in image data.

This architecture delivers best-in-class object detection efficiency and performance across a wide range of resource constraints. 
This is achieved by using EfficientNet --  state-of-the art backbone networks for feature extraction, a weighted bi-directional feature pyramid network (BiFPN), which allows easy and fast multi-scale feature fusion, and a compound scaling method that simultaneously and uniformly scales the resolution, depth, and
width for all backbone, feature, and location/class prediction networks \citep{2019arXiv191109070T}.

The use of a BiFPN, which effectively represents and processes multi-scale features, makes this architecture particularly well-suited for the problem of morphology-based comet identification and localization.

A batch of triplet image stacks of size $(n_b, 256, 256, 3)$, where $n_b$ is the number of stacks in the batch, is passed through an EfficientNet B0 backbone \citep{2019arXiv190511946T}. The extracted features from the last five blocks/levels of the network are passed through the BiFPN. The resulting five output tensors denoted in colored circles in Fig. \ref{fig:architecture} are fed into the head network, which outputs the probability of the image containing a comet $p_c$ and its centroid's predicted relative $(x, y)$ position\footnote{We note that standard object detection algorithms typically output bounding boxes and corresponding object class probabilities, i.e. sets of ($4 + n_{classes}$) numbers. Our approach allowed us to simplify the head network architecture and both simplify and speed up the assembly of the training data set, bypassing the unnecessary complexity and potential inaccuracy of drawing bounding boxes around known comet detections.}.

We defined the loss function as follows:

\begin{equation}
    L = w_c \cdot L_c + w_p \cdot L_p
\end{equation}

where $L_c$ denotes the binary cross-entropy function for the label $c$ (1 -- there is a comet in the image, 0 -- there is no comet) and the predicted probability $p_c$. 
If $\nint*{p_c} = 1$, $L_p$ is computed as an $L_1$ loss for the relative position $(x, y)$ and its prediction $(x_p, y_p)$ with a small $L_2$ regularizing term (with $\epsilon = 10^{-3}$), and $w_c$ and $w_p$ denote the weights of the two terms, respectively:

\begin{equation}
  \begin{split}
    L_c = \sum c \cdot \mathrm{log}(p_c) + (1-c) \cdot \mathrm{log}(1 - p_c) \\
    L_p = \sum \nint*{p_c} \cdot \Big(|x - x_p| + |y - y_p| + \\  \epsilon \cdot \sqrt{|x - x_p|^2 + |y - y_p|^2} \Big)
  \end{split}
\end{equation}

We employed the Adam optimizer \citep{2014arXiv1412.6980K}, a batch size of 32, and a 81\%/9\%/10\% training/validation/test data split. For data augmentation, we applied random horizontal and vertical flips of the input data; no random rotations and translations were added. We note that the test/validation sets did not contain augmented data from the training set. We used standard techniques to maximize training performance: if no improvement in validation loss was observed for 10 epochs, the learning rate was reduced by a factor of 2, and training was stopped early if no improvement was observed for 30 epochs.

The EfficientNet's weights were randomly initialized\footnote{We experimented with pre-trained weights, however that neither helped the network to reach convergence faster, nor did it affect the final performance. We believe this is likely due to the fact that astronomical images are very different from those in commonly-used data sets.}. We first set $w_c = 10, w_p = 1$ to allow for a fast convergence of the feature-extracting part of the network. To fine-tune the performance, we trained Tails on a balanced data set setting $w_c = 1.1, w_p = 1$ and monitored the validation loss for early stopping, then bumped $w_p = 2$ and monitored the validation positional RMSE; finally, added the omitted negative examples and again monitored the validation loss for early stopping.

The resulting classifiers were put through the same active-learning-like procedure as was employed in the initial data set assembly, using several months of ZTF Twilight survey data.

\section{Tails performance}

\begin{figure*}
    \centering
        \subfigure[False positive rate (FPR) and false negative rate (FNR) as a function of score $p_c$. FPR and FNR balance out at around 1.7\% for a score threshold of 0.5.]{\includegraphics[width=0.48\textwidth]{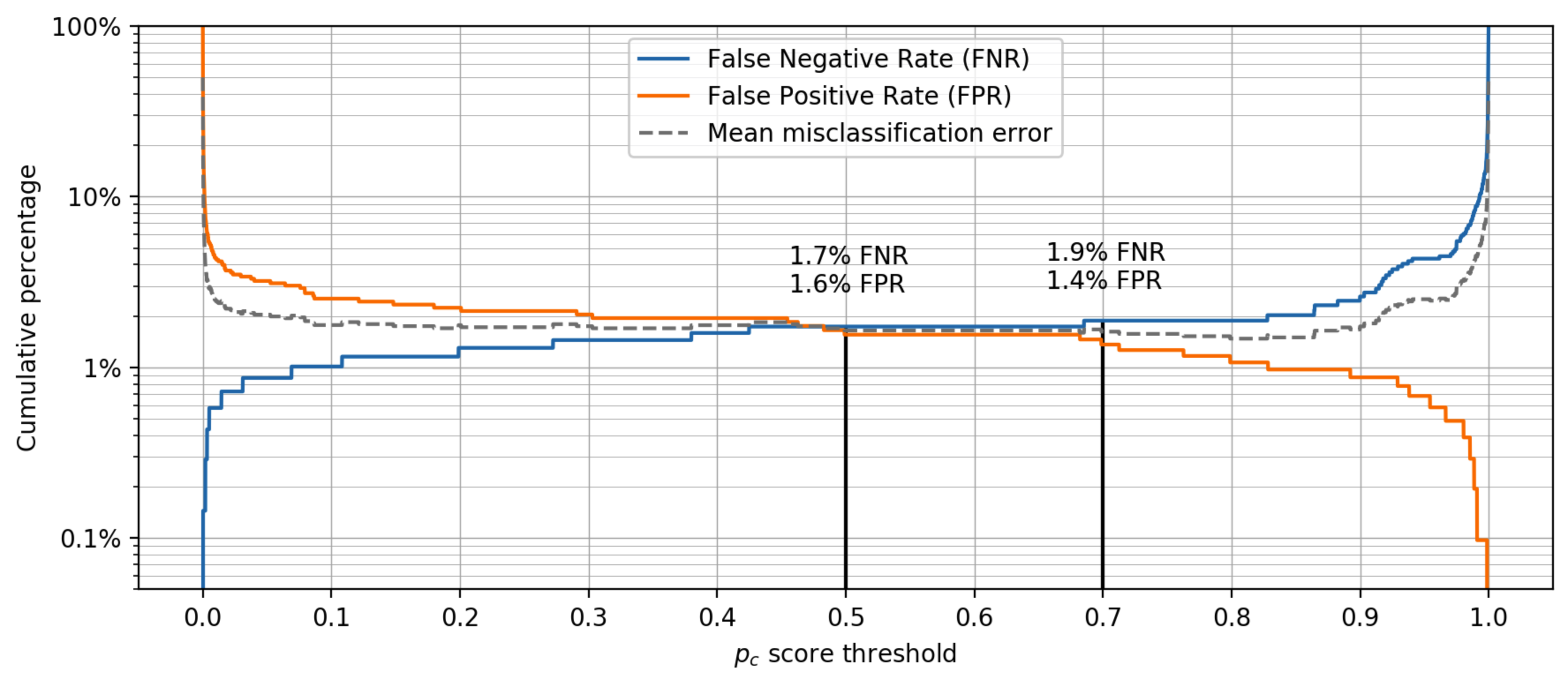}}\quad
        \subfigure[RMSE of the predicted comet position versus the reported on JPL Horizon for the 650 positive samples from the test set.]{\includegraphics[width=0.48\textwidth]{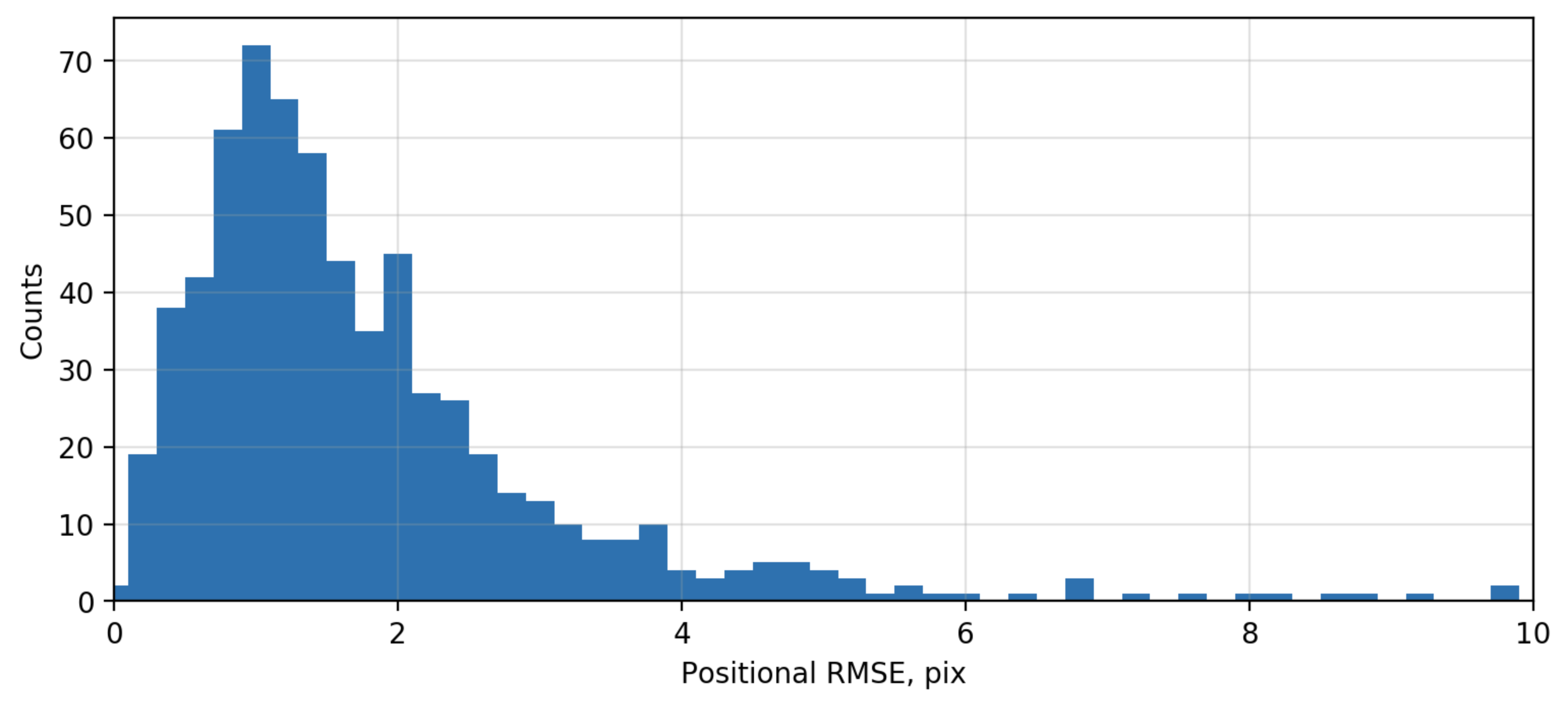}}\quad
    \caption{Test set performance of Tails. The set contains 1,400 negative and 650 positive examples.}
    \label{fig:performance}
\end{figure*}

Evaluated on the test set, with a score $p_c$ threshold of 0.5, Tails demonstrates false positive and false negative rates (FPR and FNR) of 1.7\%, and a $\sim1-2$ pixel median RMSE of the predicted comet ``centroid'' position versus that acquired from JPL Horizon (see Fig. \ref{fig:performance}).

The ZTF instrument's CCD mosaic has 16 individual \textit{6k $\times$ 6k} science CCDs. The raw ZTF image data are split
into four readout quadrants per CCD and all processing
is conducted independently on each CCD readout quadrant. We tessellate each \textit{3k $\times$ 3k} CCD-quadrant image into a 13 $\times$ 13 grid of overlapping 256 $\times$ 256 pixels tiles and evaluate Tails on those.\footnote{Standard fully-convolutional approaches often used in computer vision proved to be an overkill in this case.}

\begin{figure*}
    \centering
        \subfigure[Candidate scanning page. The users can inspect the candidates and save vetted objects to one or more groups. Candidates that are not saved to any group within 7 days are removed from Fritz.]{\includegraphics[width=0.48\textwidth]{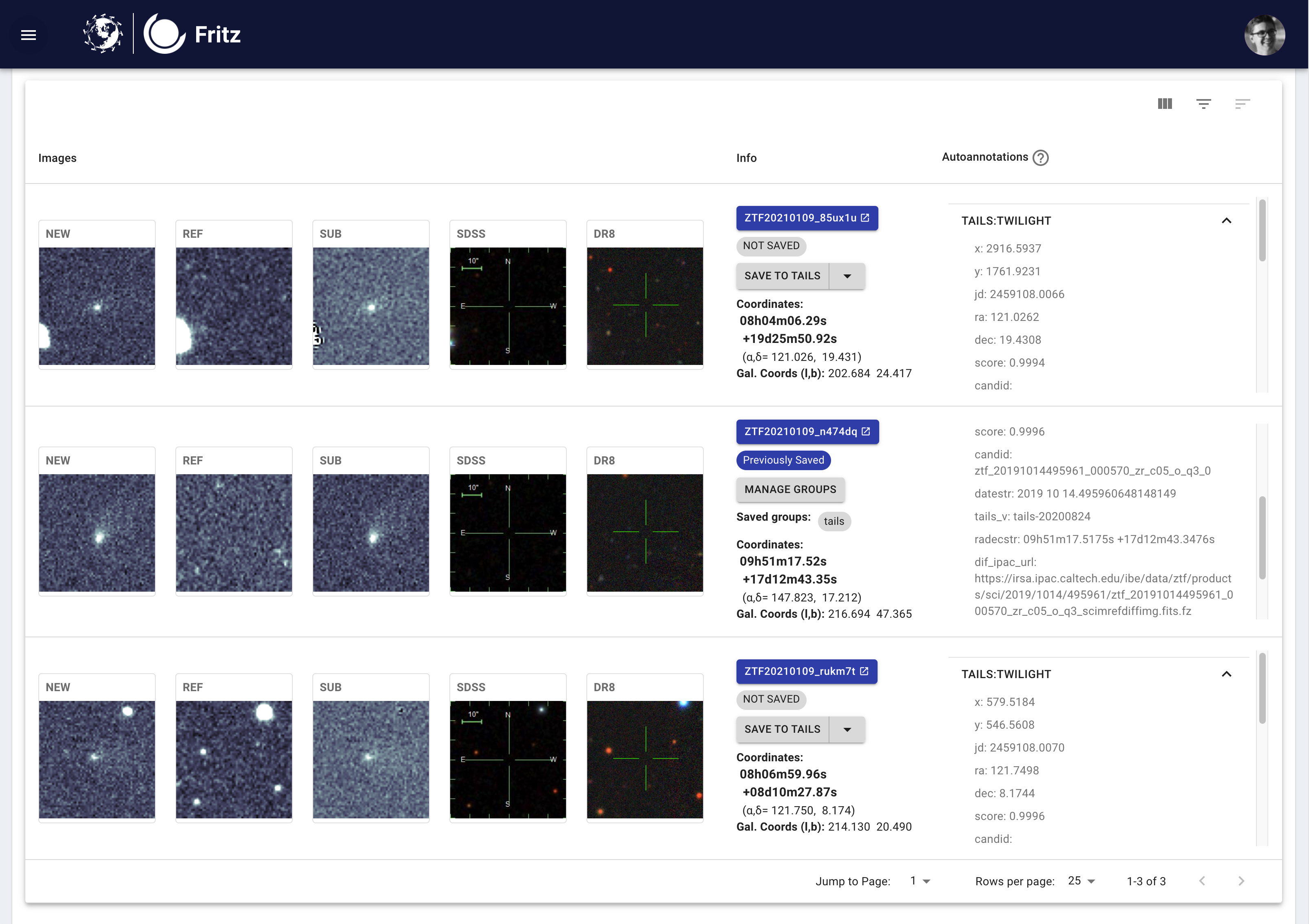}}\quad
        \subfigure[Source page. It aggregates and displays in an interactive manner all kinds of information related to an object that exists on Fritz, such as photometry, spectroscopy, auto-annotations, comments, finder charts, follow-up requests, and other data.]{\includegraphics[width=0.48\textwidth]{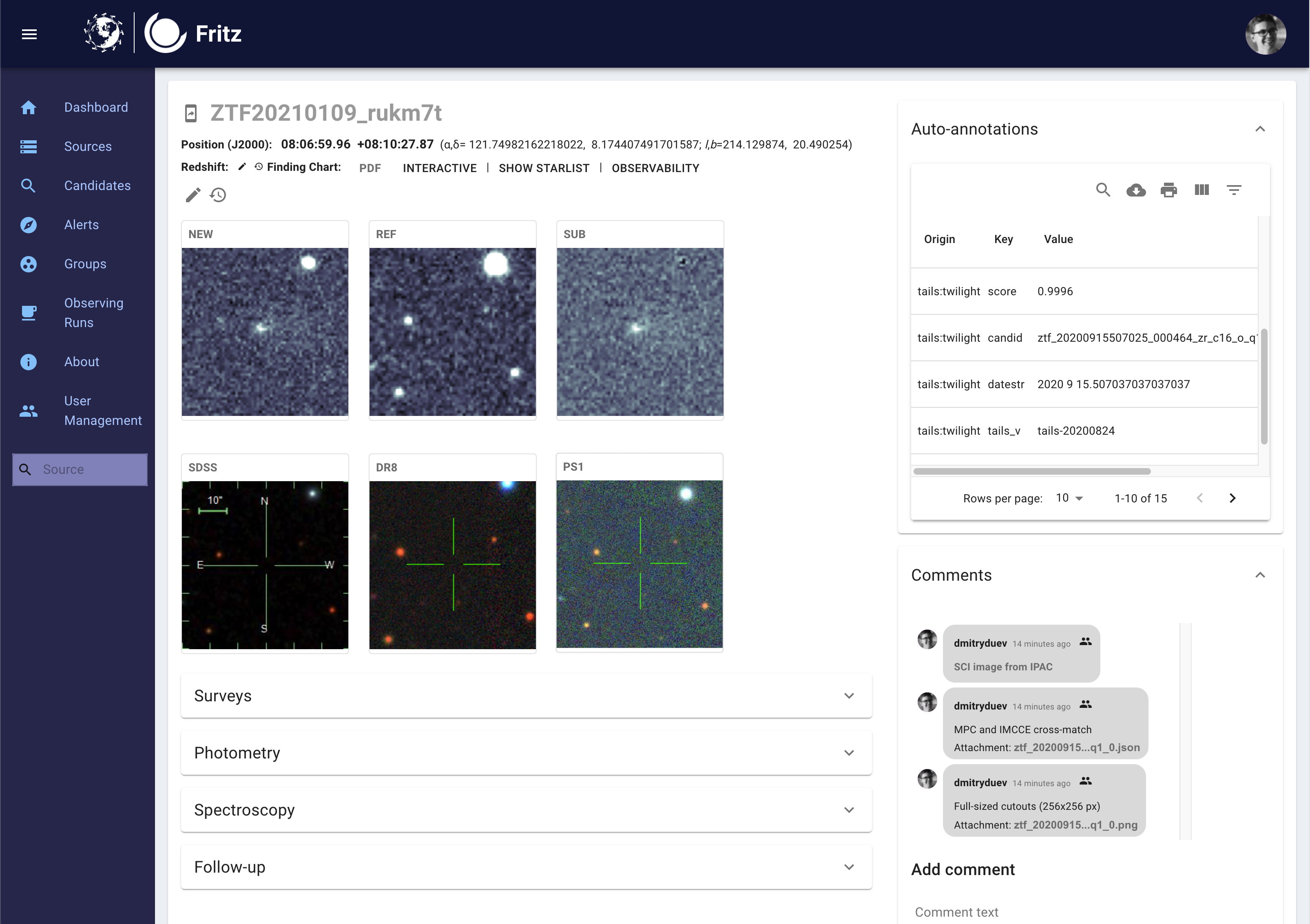}}\quad
    \caption{Screenshots of the Fritz user interfaces used for Tails candidate inspection and vetting.}
    \label{fig:fritz}
\end{figure*}

Tails has been deployed in production since late June 2020. We have implemented a ``sentinel'' service\footnote{See \url{https://github.com/dmitryduev/tails}} that processes the incoming data in real time and posts the plausible candidates to Fritz\footnote{\url{https://github.com/fritz-marshal/fritz}}, the ZTF Phase II open-source science data platform \citep{skyportal2019, 2019MNRAS.489.3582D, Kasliwal_2019}, for further manual inspection and vetting. The candidates are auto-annotated with the detailed information on the detection such as the score, CCD and sky positions, and cross-matches with known Solar System objects. Fig. \ref{fig:fritz} shows screenshots of the Fritz user interfaces used in the process.

It takes about 5 hours to run inference on a typical set of nightly ZTF Twilight data ($\sim45$ 30-second exposures) on an \textit{e2-highcpu-32} virtual machine instance (32 vCPU, 32 GB memory, SSD disk) on the Google Cloud Platform, including I/O operations.

Consistently with the expected rate of comet observations, a typical run on nightly Twilight data yields a few dozen candidates, which, given the typical number of processed tiles, gives an empirical false positive rate (FPR) value of about 0.01\%.

The scanning results are accumulated and used to expand the training set and improve Tails' performance.

We have evaluated Tails' performance on a random sample of 200 observations of known comets with identifiable morphology in July-August 2020 and found an empirical recall value of 99\%.

\begin{figure*}
    \begin{minipage}[c][11cm][t]{.5\textwidth}
      \centering
        \subfigure[2I/Borisov observed on 2019/10/05]{\includegraphics[width=1\textwidth]{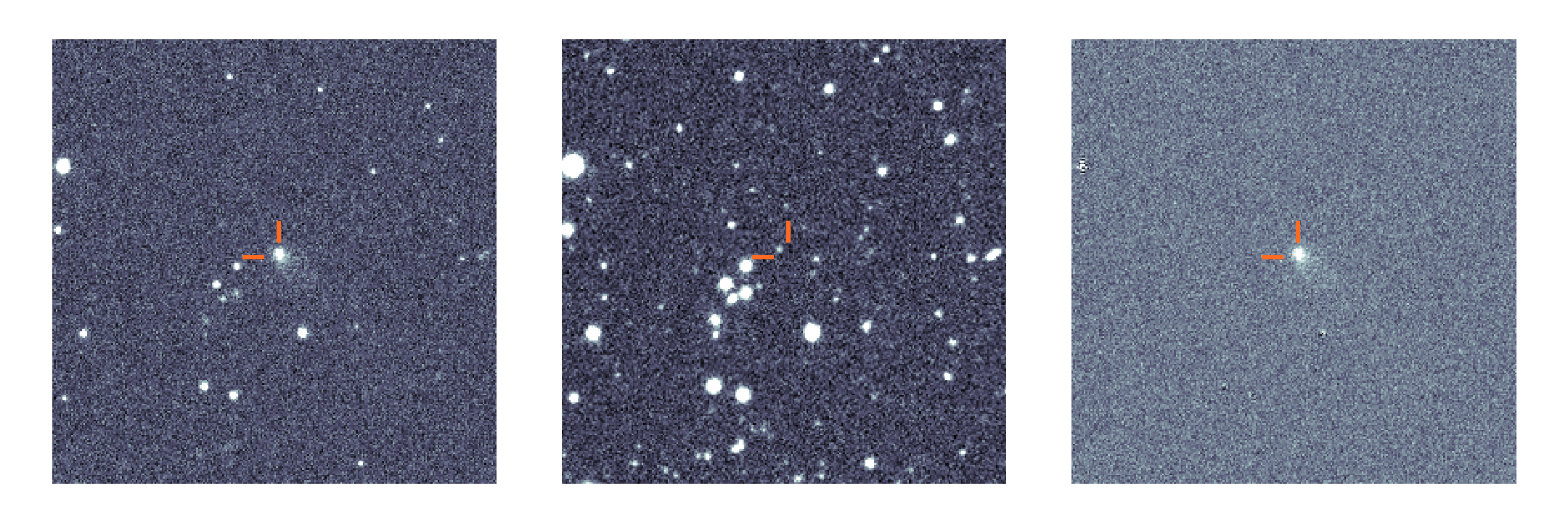}}\quad
        \subfigure[2I/Borisov observed on 2019/10/15]{\includegraphics[width=1\textwidth]{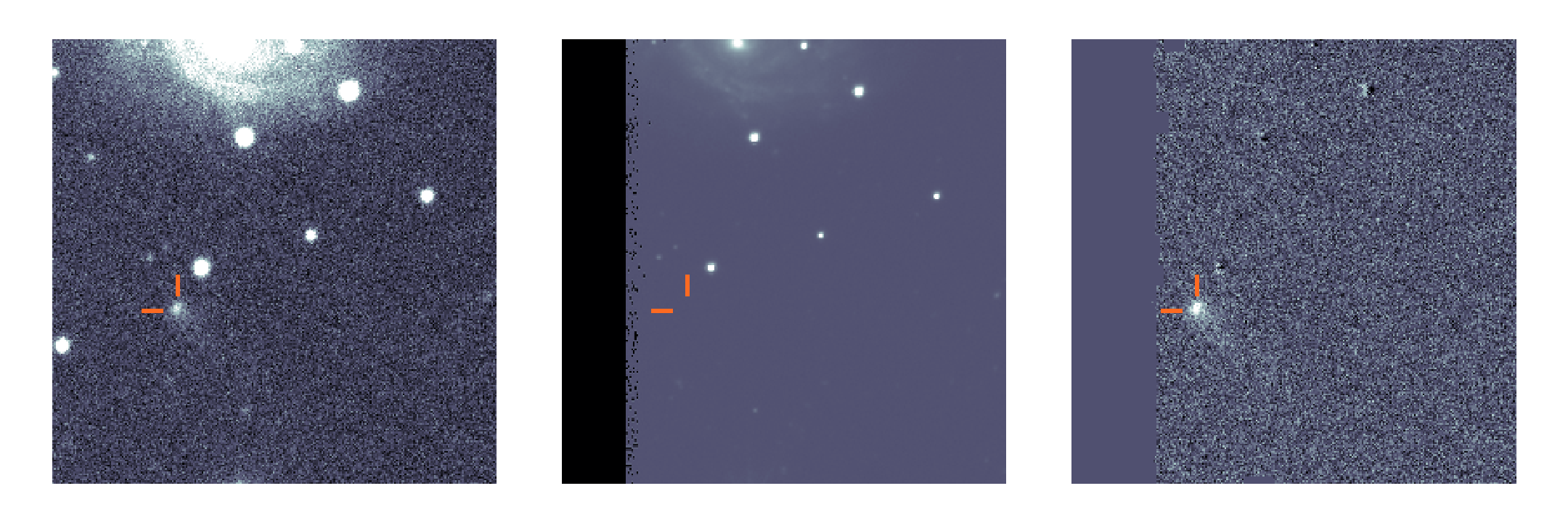}}\quad
        \subfigure[C/2010 U3 (Boattini) observed on 2020/08/11]{\includegraphics[width=1\textwidth]{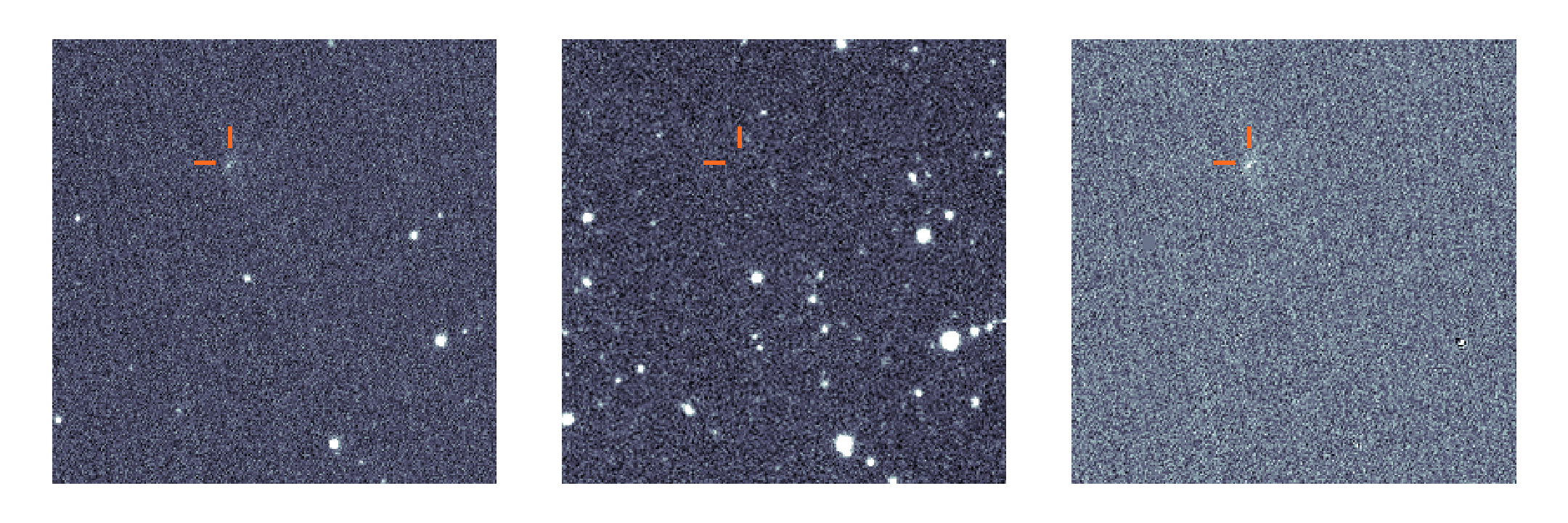}}\quad
    \end{minipage}
    \begin{minipage}[c][11cm][t]{.5\textwidth}
      \centering
        \subfigure[Bogus detection next to a bright star]{\includegraphics[width=1\textwidth]{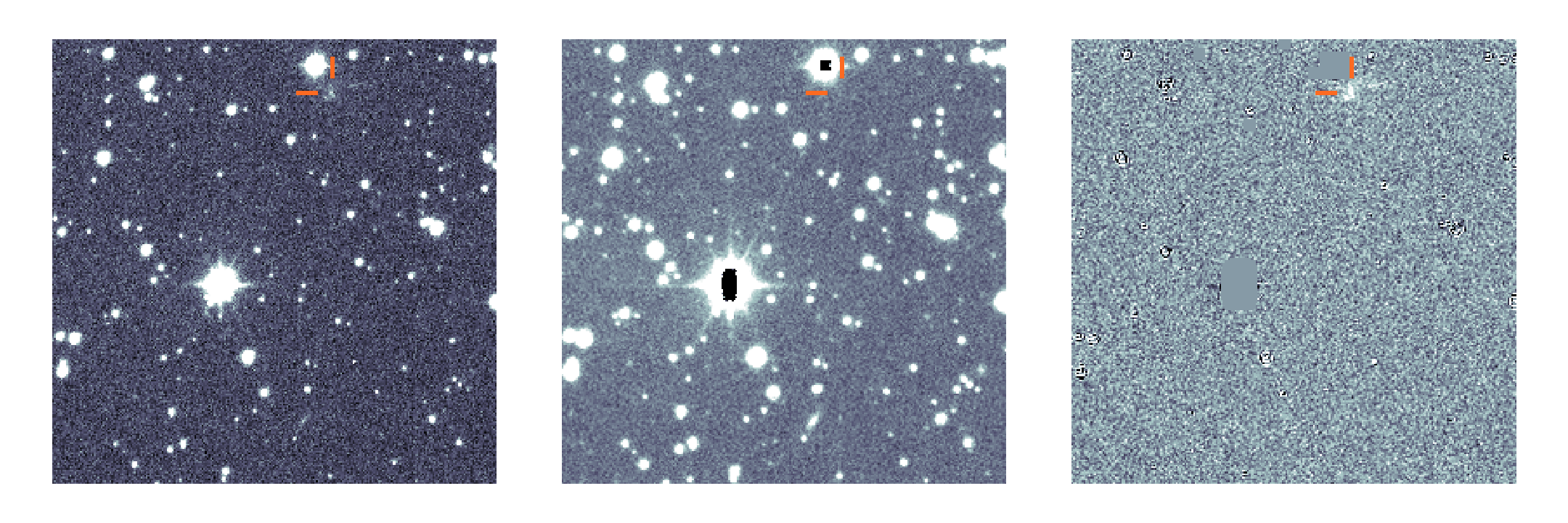}}\quad
        \subfigure[Bogus detection due to a brightened satellite trail]{\includegraphics[width=1\textwidth]{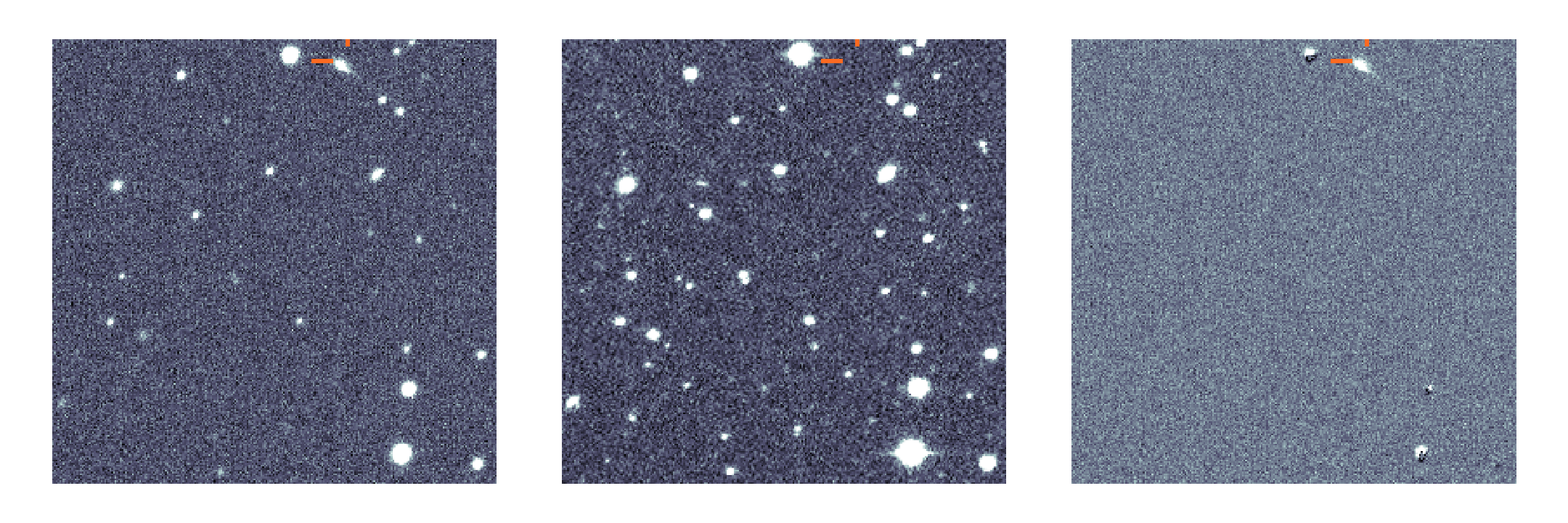}}\quad
        \subfigure[Bogus detection due to a telescope reflection]{\includegraphics[width=1\textwidth]{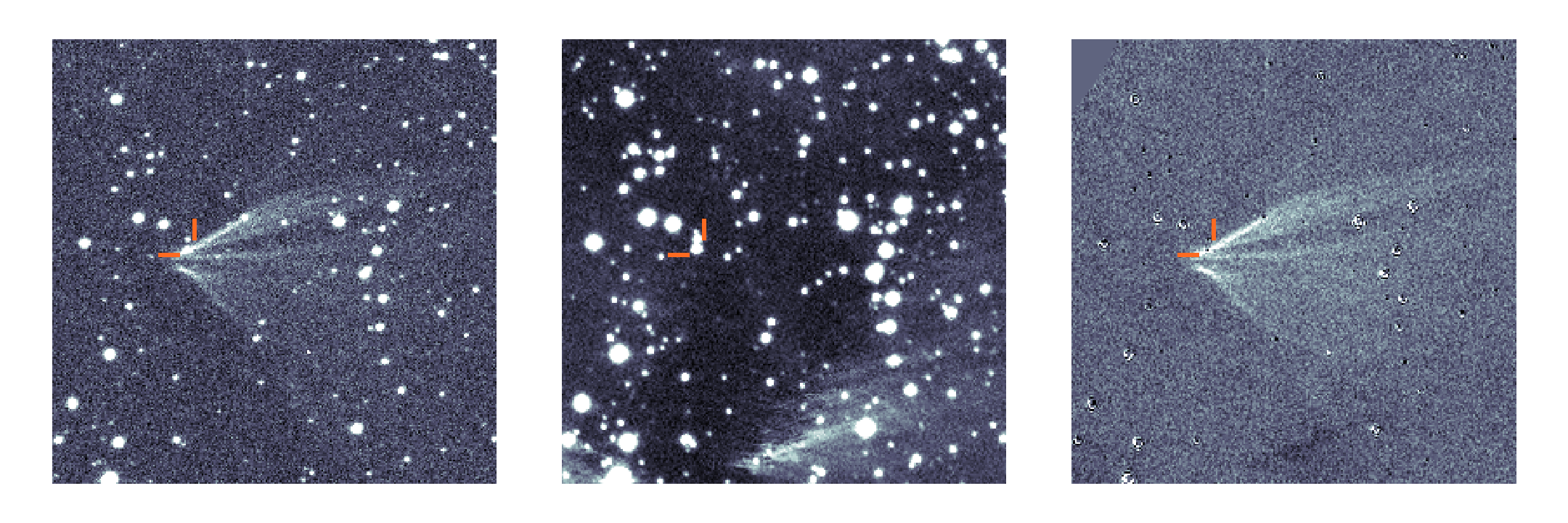}}\quad
    \end{minipage}%
    \vspace*{\fill}
    \caption{Candidates identified by Tails. Panels (a), (b), and (c) on the left show the detections of real comets. Typical false positives are shown on Panels (d), (e), and (f) on the right. 
    For each image triplet, the left pane shows the epochal science exposure, the middle pane -- the reference image of the corresponding patch of sky, and the right pane -- the ZOGY difference image.}
    \label{fig:candidates}
\end{figure*}

Fig. \ref{fig:candidates} shows a number of comet candidates not from the training set identified by Tails, including some of the ZTF observations of the comet 2I/Borisov. Optical artifacts resembling cometary objects are the main source of contamination.

\subsection{Discovery of comet C/2020 T2}

\begin{figure}
    \centering
    \includegraphics[width=0.47\textwidth]{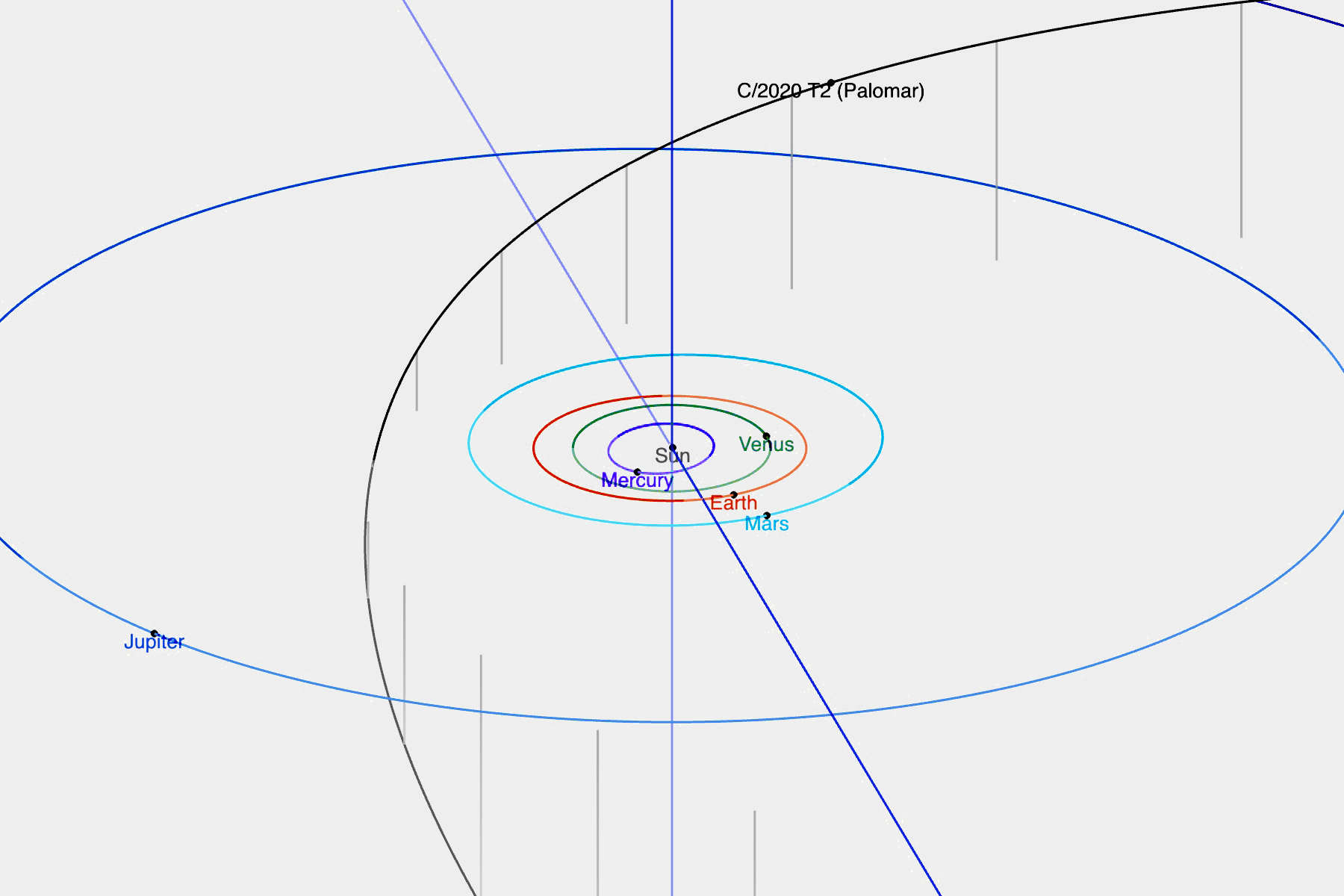}
    \caption{The orbit of comet C/2020 T2 as of October 7, 2020.
Image credit: NASA/JPL-Caltech / D. Duev.}
    \label{fig:C2020T2_orbit}
\end{figure}

\begin{figure*}
    \centering
    \includegraphics[width=0.99\textwidth]{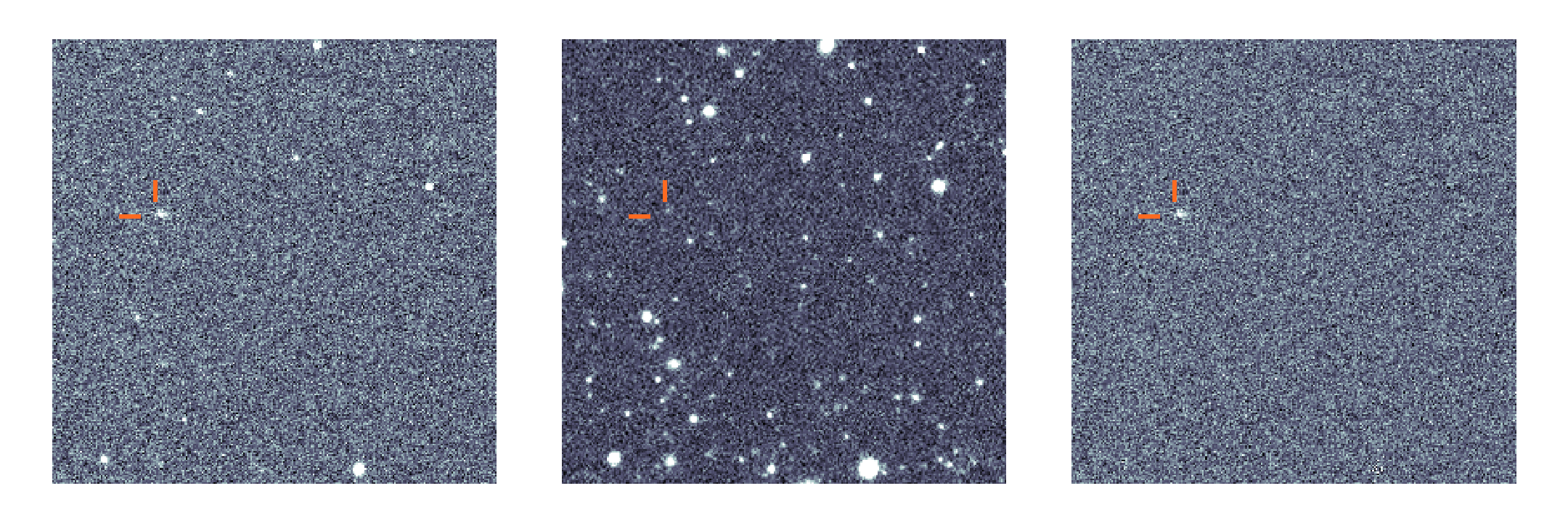}
    \caption{Discovery image of C/2020 T2 (Palomar), the first DL-assisted comet discovery by Tails. Taken on October 7, 2020 with the ZTF camera on the 48-inch Schmidt telescope at Palomar. The left pane shows the epochal science exposure (256x256 pix cutout), the middle pane -- the reference image, the right pane -- the ZOGY difference image. East is to the left, north is down.}
    \label{fig:c2020t2}
\end{figure*}

On October 7, 2020, Tails discovered a candidate that was posted to MPC's Possible Comet Confirmation Page (PCCP) \footnote{\url{https://minorplanetcenter.net/iau/NEO/pccp_tabular.html}} as ZTFDD01 (see Fig. \ref{fig:c2020t2}). It was later confirmed to be a long-period comet and designated C/2020 T2 (Palomar), marking the first DL-assisted comet discovery \citep{2020MPEC....U...170L}. The candidate was found in the Twilight survey data; it was at 19.3 mag in the ZTF $r$ band.  The FWHM of the object was approximately $2\arcsec.5$--$3\arcsec$, compared to nearby background stars that have FWHM of $\sim2\arcsec$. The object showed a tail extending up to $5\arcsec$ in the westward direction. Table \ref{tab:OrbitalElements} summarizes the orbital elements of C/2020 T2 provided by the MPC and Fig. \ref{fig:C2020T2_orbit} shows its orbit as of the discovery date.

\begin{table}[]
\begin{center}
\begin{tabular}{ll}
\hline
 element & value \\
 \hline\hline
e     & 0.9934213             \\
Incl. & 27.87307              \\
Peri. & 150.38279             \\
Node  & 83.04834              \\
q     & 2.0546940             \\
T     & 2021 July 11.14638 TT
\end{tabular}
\end{center}
\caption{Orbital elements of C/2020 T2 provided by the MPC.}
\label{tab:OrbitalElements}
\end{table}

\begin{figure}
    \centering
    \includegraphics[width=0.47\textwidth]{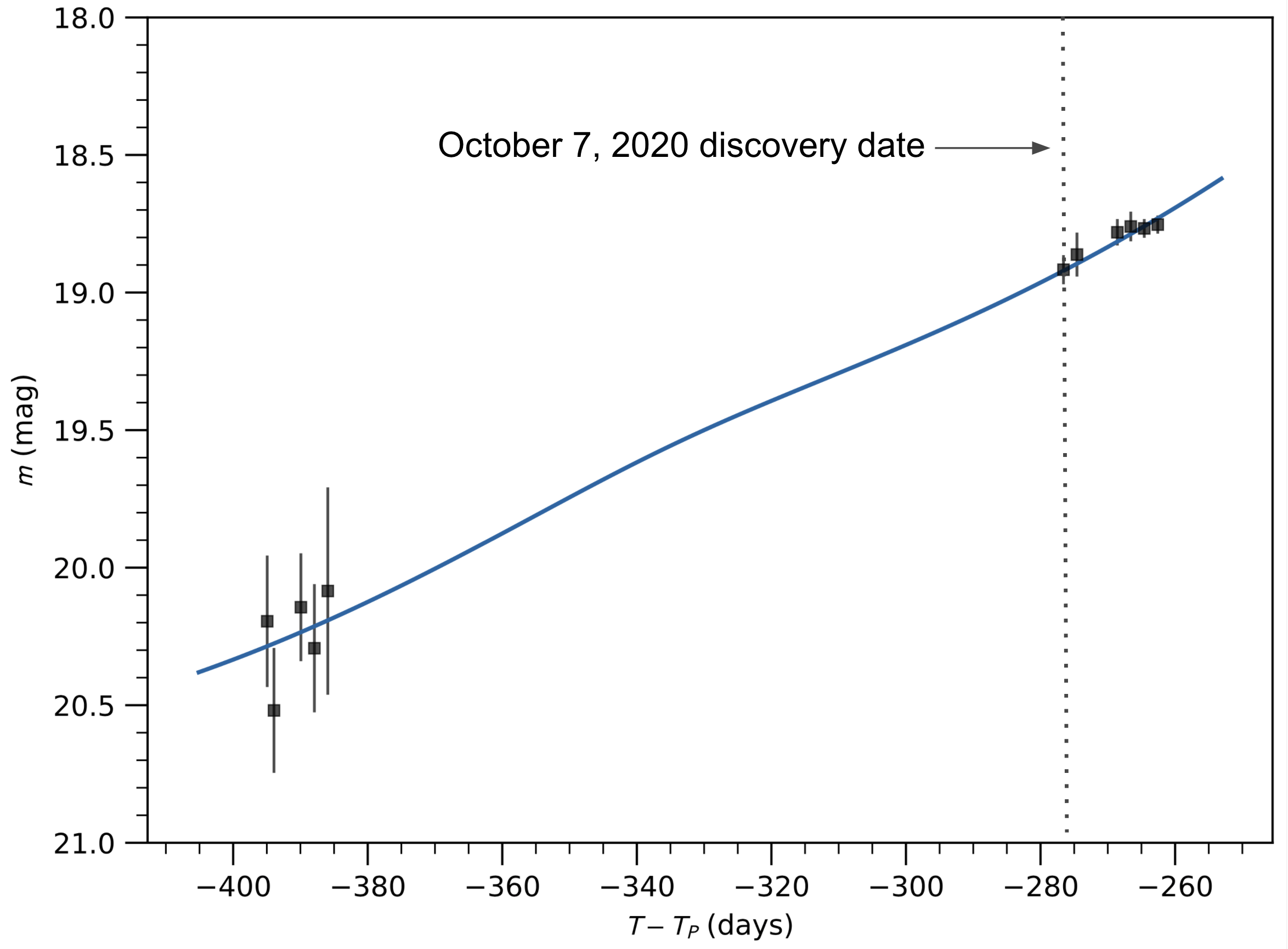}
    \caption{Photometry of comet C/2020 T2 (Palomar) derived from ZTF-TS images ($r$-band) versus time from perihelion. A best-fit model lightcurve is also shown: $r=9.85 + 9.54 \log_{10}(r_{\rm{h}}) + 5 \log_{10}(\Delta) - \Phi(\theta)$, where $r_{\rm{h}}$ is the heliocentric distance in au, $\Delta$ is the comet-observer distance in au, and $\Phi(\theta))$ is the phase angle correction from \citet{schleicher98}. $T_p$ denotes the time of perihelion passage (July 11, 2021).}
    \label{fig:lightcurve}
\end{figure}

To determine if Tails could have discovered C/2020 T2 before 2020 October 7, we searched the ZTF archive for all Twilight Survey data covering the ephemeris position of the comet with the ZChecker software \citep{2019ASPC..523..471K}.  Eleven nights of data were found between 2020 June 11--20 (evening twilight) and October 7--21 (morning twilight). The comet was in conjunction with the Sun between the two sets, and not observable by ZTF. We measured the brightness of the coma in 4-pix radius apertures, and aperture corrected the photometry according to the ZTF pipeline documentation.  The data are shown in Fig.~\ref{fig:lightcurve}. Typical seeing was $2\arcsec$ in June, the comet was very faint ($r$=20.2~mag), near the single-image detection limit ($r$=20.4--20.9~mag, 5$\sigma$ point source), and had no morphological features for Tails to pick up. Thus October 7, 2020 was really the first opportunity for Tails to discover the comet.

\subsection{Recovery of comet P/2016 J3 = P/2021 A3 (STEREO)}
\begin{figure*}
    \centering
    \includegraphics[width=0.99\textwidth]{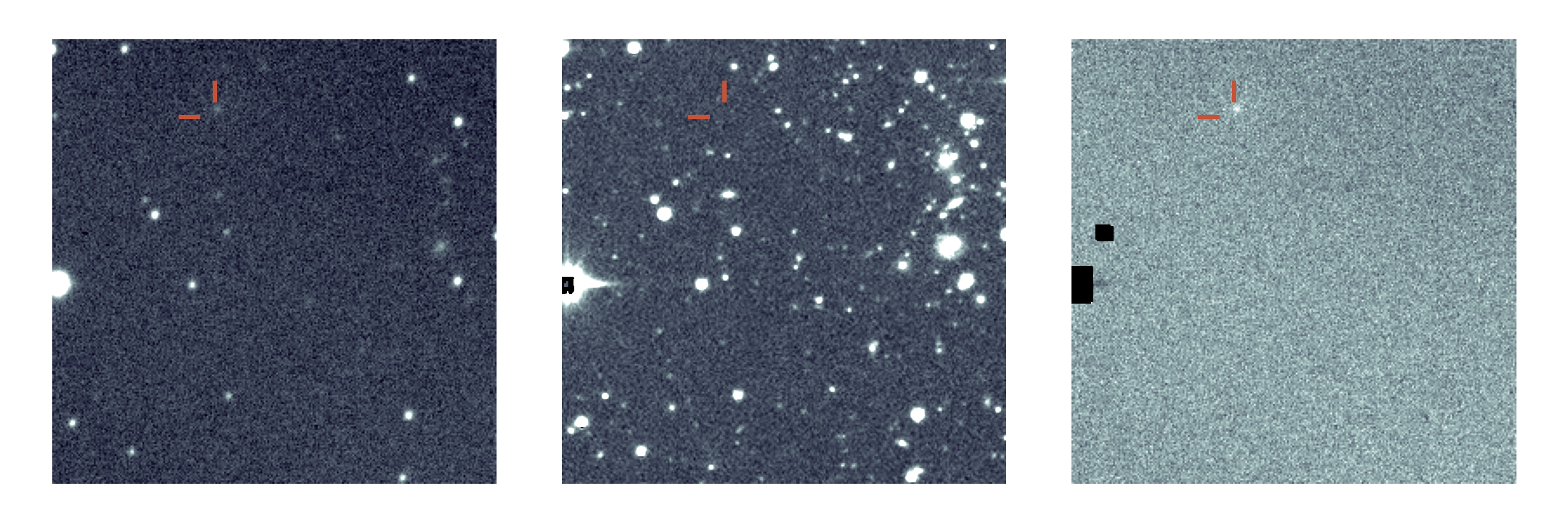}
    \caption{Recovery detection of P/2021 A3 (STEREO) identified by Tails in ZTF r-band data taken on January 4, 2021. The left pane shows the epochal science exposure (256x256 pix cutout), the middle pane -- the reference image, the right pane -- the ZOGY difference image. East is to the left, north is down.}
    \label{fig:p2021a3}
\end{figure*}

A comet candidate was identified by a combination of Tails and the ZTF Moving Object Detection Engine \citep[][]{Masci2019} on 2020 January 04 UTC and submitted to the PCCP as ZTF0Ion (see Fig. \ref{fig:p2021a3}). It was later identified as a recovery of comet P/2016 J3 (STEREO) and given the designation P/2016 J3 = P/2021 A3 (STEREO) \citep[][]{2021P3MPEC}.  P/2021 A3 was identified in the evening Twilight survey data at $r$=19.3~mag with a clearly-extended appearance scoring 0.9 with a coma $\sim$10\arcsec~wide and a tail extending past 20\arcsec in the north east direction.

\section{Discussion}

This work demonstrates the potential of the state-of-the-art deep-learning computer-vision architecture designs when applied to the problem of astronomical source detection and localization, with a specific focus on comets. 

We experimented with the input data and trained a version of Tails that instead of triplet image stacks uses duplets -- epochal/reference images, omitting the ZOGY difference images. Our tests show that this version achieves essentially the same performance as the one trained on triplets without requiring image differencing, expanding the range of potential use cases of Tails.

While Tails is trained only on ZTF data, with transfer learning, it can be adapted to other sky surveys, including the upcoming Vera Rubin Observatory's Legacy Survey of Space and Time (LSST) \citep{ivezic2008lsst}.

\acknowledgments
D. A. Duev would like to thank Ivan Duev for assistance with data labeling. D. A. Duev acknowledges support from Google Cloud and from the Heising-Simons Foundation under Grant No. 12540303.

Based on observations obtained with the Samuel Oschin Telescope 48-inch and the 60-inch Telescope at the Palomar Observatory as part of the Zwicky Transient Facility project. ZTF is supported by the National Science Foundation under Grant No. AST-1440341 and a collaboration including Caltech, IPAC, the Weizmann Institute for Science, the Oskar Klein Center at Stockholm University, the University of Maryland, the University of Washington, Deutsches Elektronen-Synchrotron and Humboldt University, Los Alamos National Laboratories, the TANGO Consortium of Taiwan, the University of Wisconsin at Milwaukee, and Lawrence Berkeley National Laboratories. Operations are conducted by COO, IPAC, and UW.

This research has made use of data and/or services provided by the International Astronomical Union's Minor Planet Center.

The authors would like to express gratitude to the anonymous referee.

\vspace{5mm}
\facilities{PO:1.2m, ZTF}

\software{astropy \citep{2018AJ....156..123A},
          Fritz (\url{https://github.com/fritz-marshal/fritz}),
          Kowalski \citep{2019MNRAS.489.3582D},
          matplotlib \citep{Hunter:2007},
          numpy \citep{2020arXiv200610256H},
          pandas \citep{reback2020pandas},
          pypride \citep{2016A&A...593A..34D},
          SEP \citep{barbary16-sep},
          sbpy \citep{2019JOSS....4.1426M},
          TensorFlow \citep{2016arXiv160304467A},
          ZChecker \citep{2019ASPC..523..471K}
          }

\bibliography{tails}{}
\bibliographystyle{aasjournal}

\end{document}